\documentclass[aps,amssymb,amsmath, twocolumn, pra, superscriptaddress]{revtex4}
\usepackage{graphicx}
\usepackage{epsfig}
\usepackage{dcolumn}
\usepackage[up]{subfigure}
\usepackage{float}
\usepackage{dsfont}	
\usepackage{relsize}
\newcommand{\tr}{\textcolor{red}}

\usepackage{epstopdf}

\usepackage{float}
\usepackage{ragged2e}
\usepackage{braket}
\usepackage[font=small,labelfont=bf,justification=raggedright, format=plain]{caption}
\usepackage{setspace}
\usepackage[colorlinks, citecolor = magenta]{hyperref}
\makeatletter
\begin{document}
	\title{Second-order correlations and purity of unheralded single photons from \\ spontaneous parametric down-conversion}
	\author{A. Padhye}
	\email{anujapadhye@iisc.ac.in}
	\affiliation{Quantum Optics \& Quantum Information, Department of Instrumentation and Applied Physics, Indian Institute of Science, Bengaluru 560012, India}
	\author{ K. Muhammed Shafi}
	\email{muhammeds@iisc.ac.in}
	\affiliation{Quantum Optics \& Quantum Information, Department of Instrumentation and Applied Physics, Indian Institute of Science, Bengaluru 560012, India}
	\author{C. M. Chandrashekar}
	\email{chandracm@iisc.ac.in}
	\affiliation{Quantum Optics \& Quantum Information, Department of Instrumentation and Applied Physics, Indian Institute of Science, Bengaluru 560012, India}
	\affiliation{The Institute of Mathematical Sciences, C. I. T. Campus, Taramani, Chennai 600113, India}
	\affiliation{Homi Bhabha National Institute, Training School Complex, Anushakti Nagar, Mumbai 400094, India}

	\begin{abstract}
		Various quantum technology applications require high-purity single photons with high generation rate. Although different methods are employed to generate such photons,  heralded single photons from spontaneous parametric down-conversion (SPDC) is the most commonly used approach. Photon generation rate from the heralded single-photon sources are limited by the efficiency of the detectors to record  coincidence detection of the photon pairs which are lower than the single-photon counts recorded separately on each detector. In this paper we present a revised expression to calculate second-order temporal correlation function, $g^{(2)}$ for any fixed time window (bin) and report the experimental characterization of purity of unheralded and heralded single photons from the SPDC process. With an appropriate choice of time bin for a given pump power, without heralding we show that higher rate of single photons with $g^{(2)}(0) = 0$ can be generated with very high probability. 
	\end{abstract}
	
	\maketitle
	
	
	\section{\label{sec1}Introduction}

Single photons and entangled photon pairs are central to the field of modern quantum optics and have enabled major advances in photonic quantum information processing tasks\,\cite{BPM97, MWK96, JSW00, PBW98, KLM01, MZK12}. The process of spontaneous parametric down-conversion (SPDC) is among the most matured and widely used approach to generate such photons\,\cite{APS21,KE07}. In SPDC process, a pump photon interacts with an optically nonlinear medium with very low probability and splits into a pair of photons with lower energy than that of the pump photon. Detection of such photons at lower energy signals the presence of its partner photon by virtue of conservation of energy, and conservation of momentum signals the direction of the arrival of the photon pair. By carefully engineering the SPDC process, the photon pair can also be ensured to be entangled in the polarization degree of freedom. Although SPDC process is probabilistic in nature, its compactness, long coherence times, and ease of operation at room temperature has made it a standard source of single photons and entangled photons\,\cite{KMW95,KFM04,FKW05}. 
	
Since SPDC process produces photons in pairs, presence of single photons can be gated by heralding one of the photons\,\cite{FAT04,BCR09}. The photon generation rate from these heralded single-photon sources is dependent on the efficiency of coincidence detection of the photon pairs and it is lower than the photon counts recorded independently due to lack of highly efficient single-photon detectors\,\cite{EFM11,RHH09}. If we are not gating the presence of photon by heralding,  the field striking on the detector is generally treated as a thermal source when one examines its full temporal behaviour. Instead of examining the full temporal behaviour, if we characterize the field in smaller time windows (bins) we can determine the presence of single photons. It will depend on the ability of the experimental setup to resolve the photon arrival time together with low dead time of the detectors. Since SPDC process is highly probabilistic in nature, even when we choose a very small time bin, along with bins containing single photons\,\cite{AC2012}, we will also have bins with multi-photons due to photon bunching effect.  Detection of multi-photons leads to compromise on the purity of single photons making it unreliable for applications which require single-photon state of high purity.  In the interest of wide application, methods to improve purity of single photons with high generation rate is one of the most pursued areas of research\,\cite{EM2020,PM2008,AZ2011,WT2017,SKB09, MMM19}. 
	
Measurement of second-order temporal correlation, $g^{(2)}(\tau)$ reveals unique statistics of the light source\,\cite{BT56,GRA86}. In particular, function $g^{(2)}(\tau )$ for a single-photon source is a measure of the likelihood of observing another photon in certain time interval $\tau$ after an initial photon is detected. Therefore, $g^{(2)}(\tau )$ looks at the difference in arrival time between consecutive photons and it is used for quantifying the purity of single photons. Signature of an ideal single-photon source is $g^{(2)}(0) = 0$ and $g^{(2)}(0) \ge 1$ indicates a classical field\,\cite{MW1994}. 
	
In this paper, we calculate $g^{(2)}(\tau)$ using probabilities of detection in each detector for a fixed time bin rather than usually followed method of determining probabilities from the full temporal behaviour and averaging for the time bin\,\cite{BCR09}. By calculating  $g^{(2)}(\tau)$  as a function of time bin $\tau$ for different pump power, we show that the value of  $\tau$ for which $g^{(2)}(\tau) = 0$ gets narrower with increase in pump power and becomes greater than 0 for very high pump power.  By analyzing the recorded photons across large number of bins with different bin width, we show the change in number of bins containing no-photon, single-photons, and multi-photons with increase in $\tau$.  This characterization helps us to associate the combination of time bin and pump power with the purity of single photons and choose the appropriate time bin and pump power accordingly to keep the multi-photon noise to the minimum. Our study also shows that for a given pump power with smaller time bin, unheralded source results in higher single-photon counts and low multi-photon noise when compared to heralded source.  Thus, with a choice of low pump power, better detector efficiency with low dead time, and filtering of pump power from entering detector can ensure unheralded single photons with high purity from SPDC process and can be effectively used for applications where gated or on-demand sources of single photons are not essential\,\cite{SGA21, SCH22}. 
	
The paper is organized as follows. In section\,\ref{sec2}, we present a formalism to calculate the second-order correlation function for a fixed time bin under unheralded and heralded conditions. In section\,\ref{Exptsch} we present an experimental setup, the measurement procedure and the results showing $g^{(2)}(\tau)$ value as a function of time bin measured for different pump power. We summarize the results and conclude with remarks in section\,\ref{conc}.

	
	\section{Second-order correlation function for a fixed time bin} 
	\label{sec2}
	
	The statistical properties of the single-photon source can be analyzed in terms of the second-order correlation function. Hanbury Brown–Twiss (HBT) setup\,\cite{BT56} is one of the well established methods to measure $g^{(2)}(0)$, by introducing a 50:50 beam splitter in the path of the light and measuring the outputs of the beam splitter using two detectors,
	\begin{equation}
		g_{\mathrm{2d}}^{(2)}(0) = \frac{P_{AB}}{P_{A} P_{B}}.
	\end{equation}
	In the preceding expression, $P_{A}(P_{B})$ is the probability of detection at detector $A$ ($B$) and $P_{AB}$ is the probability of twofold coincidence detection in detectors $A$ and $B$ in a short instance of time.  For SPDC process which generates pair of down-converted photons, same approach of two detector measurement of $g^{(2)}$ is followed when they are unheralded, by sending one of the SPDC photons through the beam splitter and not heralding its partner photons.  When single photons from SPDC process are heralded  using detector C, a three detector measurement procedure is followed, 
	\begin{equation}
		g_{\mathrm{3d}}^{(2)}(0) = \frac{P_{ABC}}{P_{AC} P_{BC}}.
	\end{equation}
	Here,  $P_{AC}$ ($P_{BC}$) is the probability of joint detection in detectors $A$ and $C$ ($B$ and $C$) and  $P_{ABC}$ is the probability of threefold coincidence detection between all the detectors in short time interval. These probabilities of detection to obtain $g^{(2)}$ value are usually calculated by averaging the counts from the total counting time\,\cite{Bec07, JFC18}.  When single photons are used for quantum technology applications or as a source for quantum optics experiments, states of each photons are measured along with their arrival time, therefore, characterizing $g^{(2)}$ for each time bin and averaging will be more relevant.  Therefore, rather than considering the full temporal behaviour, we will focus on probabilities of events within the fixed time bin $\tau$  and calculate  averaged $g^{(2)}(\tau)$ value for large sample of such bins.  For unheralded photon source,  
	\begin{equation}
		g_{\mathrm{uh}}^{(2)}(\tau) = \frac{1}{N_w(\tau)} \sum_{N_w}  \frac{P_{AB}(\tau)}{P_{A}(\tau)P_{B}(\tau)},
	\end{equation}
	where $P_{A}(\tau)$,  $P_{B}(\tau)$, and $P_{AB}(\tau)$ are the probability of detection in detector A,  detector B, and simultaneous detection in both the detectors in the time bin $\tau$, respectively.  Since SPDC process is probabilistic in nature, for low pump power or when bin size is very small we can have bins with no-photon detection and value of $g^{(2)}(\tau)$ for such bins will be undefined. Therefore, the average is taken only over the number of bins with at least one-photon detection. Thus, the preceding expression is averaged over total number of bins with at least one-photon detection, $N_w$. Where, $C_{A}(\tau)$ and $C_{B}(\tau)$ are the number of photons detected at detectors $A$ and $B$ respectively in the time bin $\tau$. The total number of counts recorded at the two detectors will be $C_{A} (\tau)+ C_{B}(\tau)$. Hence,  
	\begin{align}
		P_{A}(\tau) = \frac{C_{A}(\tau)}{C_{A}(\tau) + C_{B}(\tau)}, \nonumber \\
		P_{B}(\tau) = \frac{C_{B}(\tau)}{C_{A}(\tau) + C_{B}(\tau)}, \nonumber \\
		P_{AB}(\tau) = \frac{C_{AB}(\tau)}{C_{A}(\tau) + C_{B}(\tau)}.
	\end{align}
	Here, the coincidence count of photons in the time bin $\tau$ is $C_{AB}(\tau) = \mathrm{min}\{C_{A}(\tau), C_{B}(\tau)\}$.  Thus, 
	\begin{align}
		g_{\mathrm{uh}}^{(2)}(\tau) = \frac{1}{N_w(\tau)} \sum_{N_w} \frac{C_{AB}(\tau)\big (C_{A}(\tau) + C_{B}(\tau)\big )}{C_{A}(\tau)C_{B}(\tau)}.  
		\label{g2uherald}
	\end{align}
	Similarly, for a heralded photon source, 
	\begin{align}
		g_{\mathrm{h}}^{(2)}(\tau) = \frac{1}{N_w(\tau)} \sum_{N_w} \frac{C_{ABC}(\tau) C_{C}(\tau) }{C_{AC}(\tau)C_{BC}(\tau)},  
		\label{g2herald}
	\end{align}
	where, $C_{C}(\tau)$  is the number of photons detected in detector $C$ in the time bin $\tau$ and  $C_{ABC}(\tau) = \mathrm{min}\{C_{A}(\tau), C_{B}(\tau),  C_{C}(\tau) \}$,  $C_{AC}(\tau) = \mathrm{min}\{C_{A}(\tau), C_{C}(\tau)\}$ and  $C_{BC}(\tau) = \mathrm{min}\{C_{B}(\tau), C_{C}(\tau)\}$.

	\section{Experimental Method}
	\label{Exptsch}
\begin{figure}[h!]
		\centering
		\includegraphics[width=0.48\textwidth]{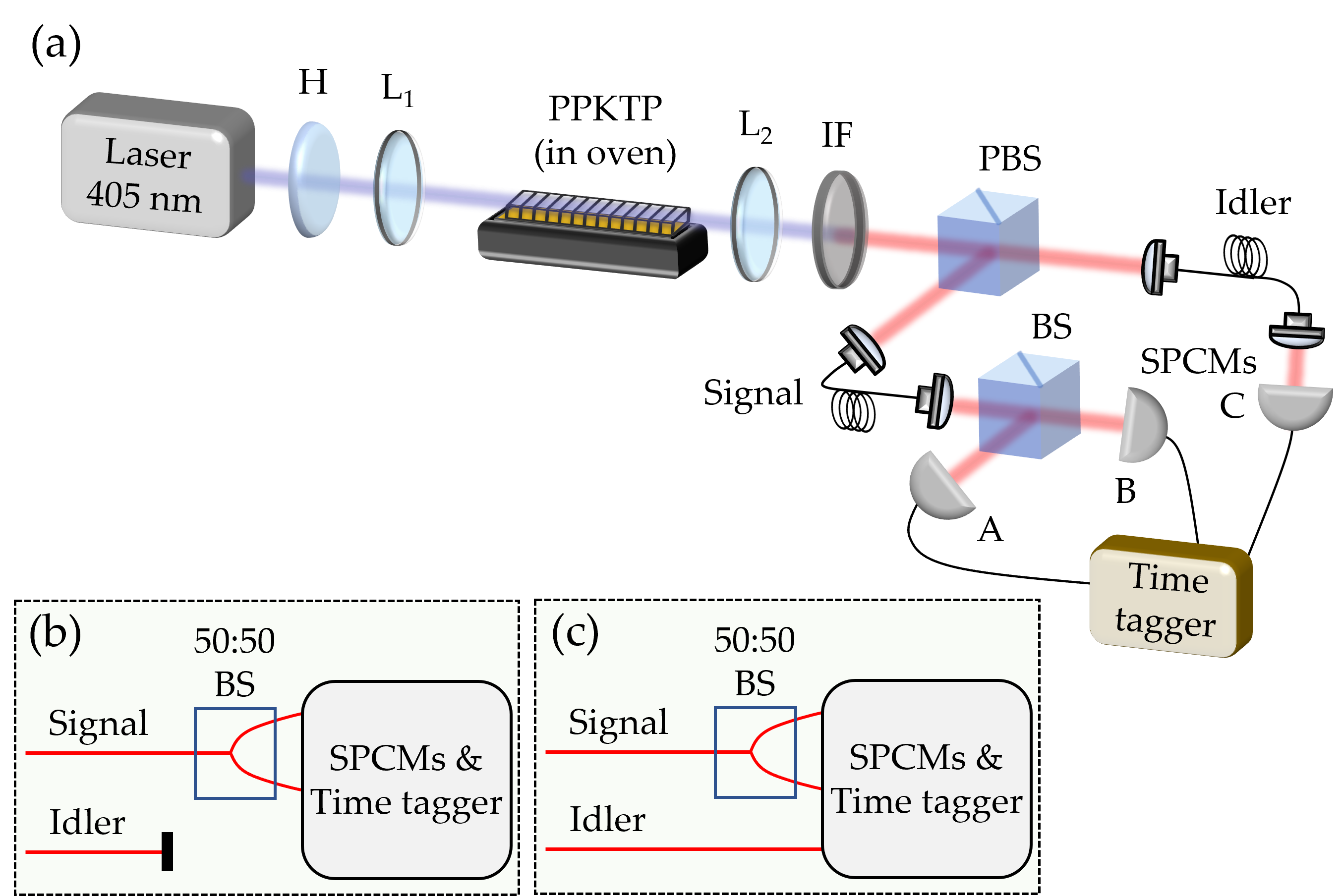}
		\caption{(a) Experimental setup: nonlinear crystal PPKTP is pumped by 405 nm laser. Half wave plate (H) is used to control the pump polarization and lens (L$_{1}$ of 300 mm) is used to focus the laser beam into the center of the crystal. Lens (L$_{2}$ of 35 mm) is used to collimate the down-converted photons. Bandpass interference filter (IF) at 810 nm $\pm$ 10 nm is used for filtering the pump. A polarization beam splitter (PBS) is used to separate the idler and signal photons. Non-polarizing 50:50 beam splitter (BS) is placed in the signal arm and SPCMs (A \& B) are used to detect single photons at the output of BS. When idler is used for heralding, it is detected through SPCM (C).  The output of the SPCMs are fed to a time-correlated single photon counter (Time Tagger).  (b) Configuration for unheralded $g_{\mathrm{uh}}^{(2)}(\tau)$ measurement (c) Configuration Heralded $g_{\mathrm{h}}^{(2)}(\tau)$ source.}
		\label{Fig1}
	\end{figure}	
\begin{figure}[h!]
		\centering
		\begin{subfigure}[ ]{ 
			\includegraphics[width=0.46\textwidth]{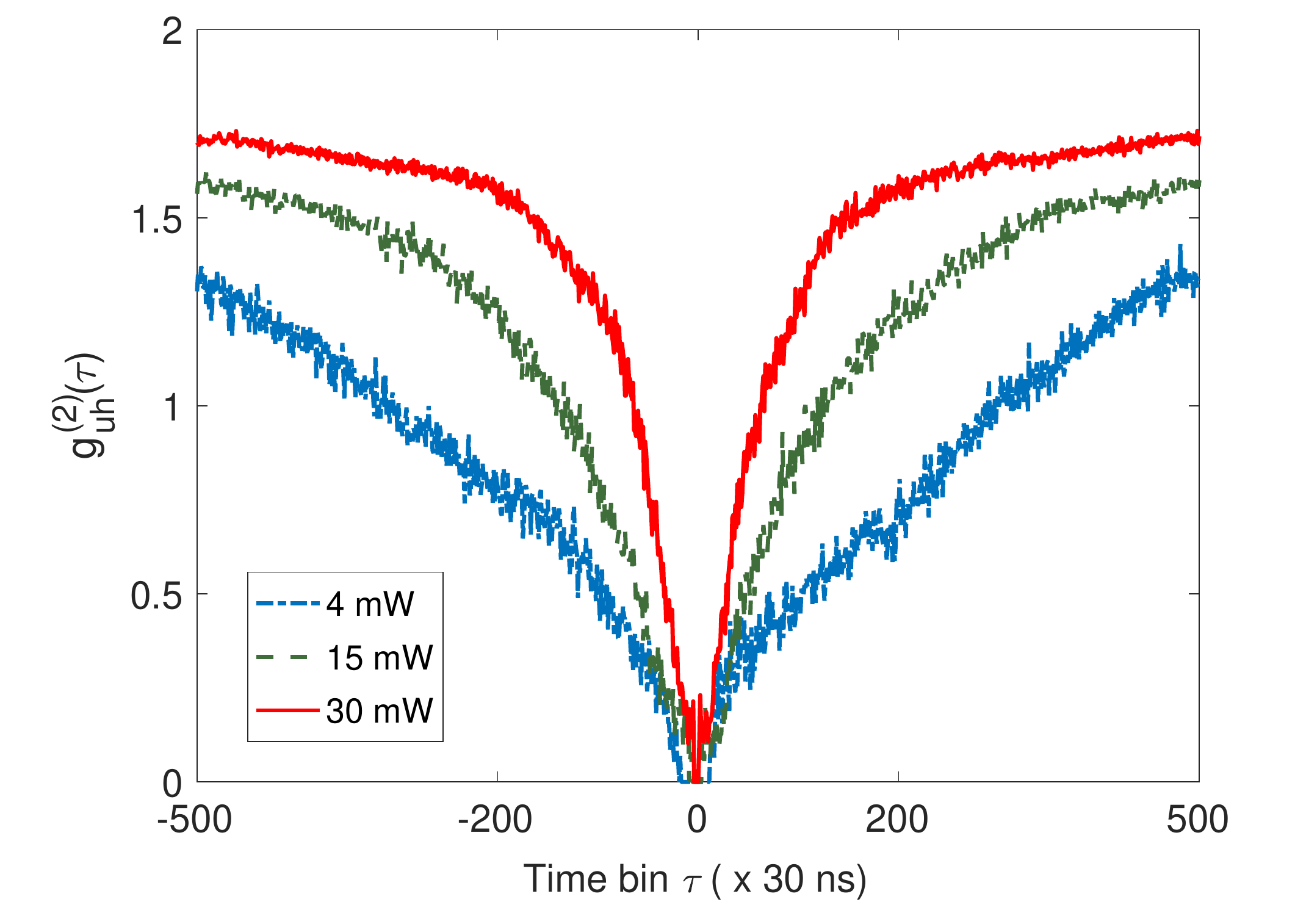}
			\label{2a}}
	 \end{subfigure}
		\begin{subfigure}[ ]{ 
		\includegraphics[width=0.46\textwidth]{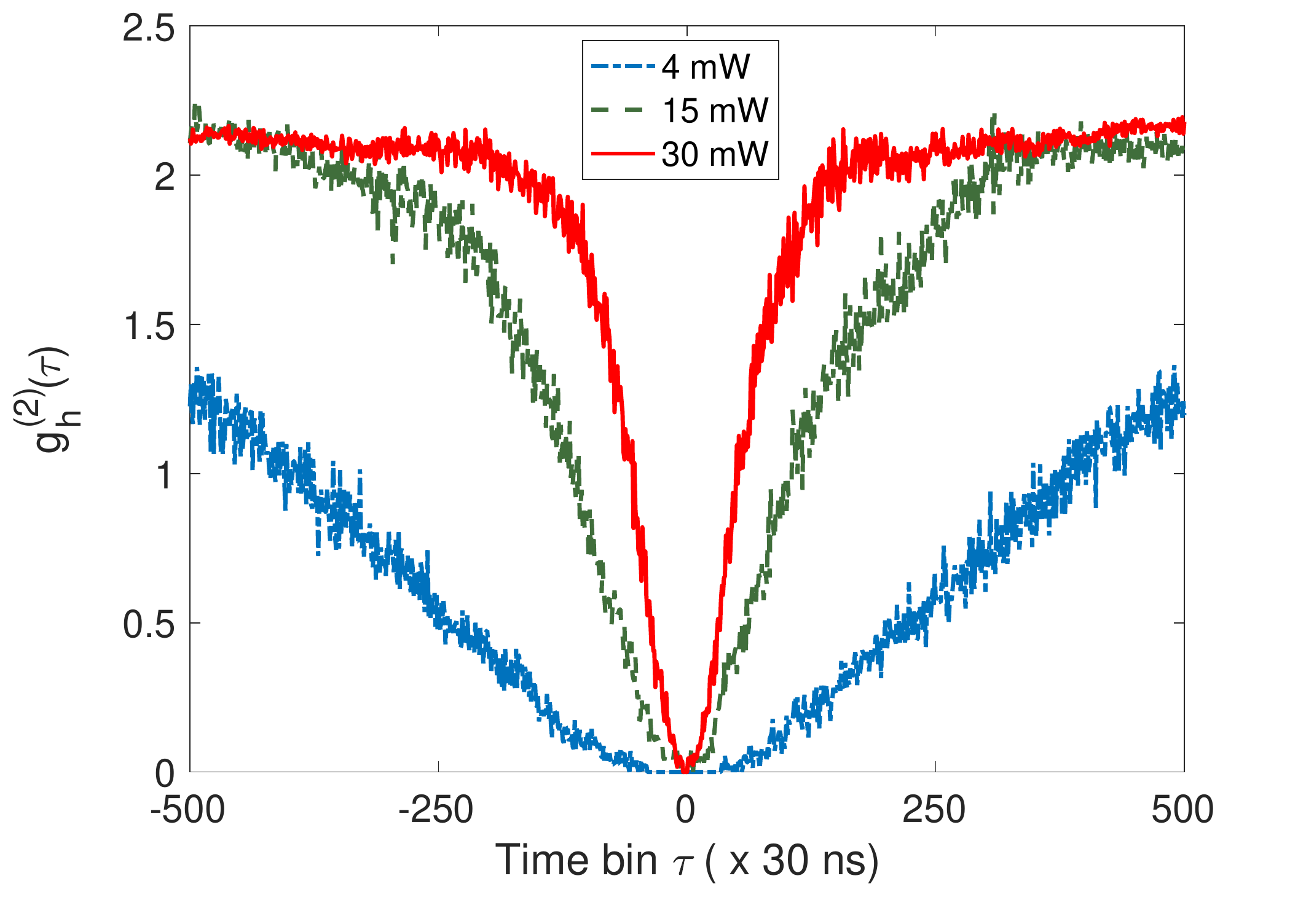}
		  \label{2b}}
        \end{subfigure}
		\caption{Variation of $g^{(2)}(\tau)$ with increase in time bin $\tau$ along both the directions from the fixed point in the photon detection time. Minimum $\tau$ was set to 30 ns $>$ 22 ns (detector dead time). Function $g^{(2)}(\tau)$ is calculated, for photon counts recorded using single-mode optical fiber connections to the detectors, for 200 samples for each $\tau$ and is averaged over the total number of bins with at least one-photon detection. We have calculated $g^{(2)}(\tau)$ value for three different pump power. (a) $g_{\mathrm{uh}}^{(2)}(\tau)$ calculated for the unheralded setup. (b) $g_{\mathrm{h}}^{(2)}(\tau)$ calculated for the heralded source. For both (a) and (b), we can see a clear dip at zero time bin leading to  $g_{\mathrm{h}}^{(2)}(0) = 0$ indicating the purity of single photons. With increase in pump power the time bin containing single photons decreases.}  
		\label{g2UnHandH}
	\end{figure}
	\subsection{Experimental setup}
	The purity of single photons generated from SPDC process is measured using the second-order correlation function, $g^{(2)}(\tau)$. Fig.\,\ref{Fig1} shows the schematic of the experimental setup to generate single photons and collect data to calculate $g^{(2)}(\tau)$. Type-II SPDC process is deployed for generating single photons using a 10-mm-long PPKTP (periodically-poled potassium titanyl phosphate) nonlinear crystal (Raicol) with poling period $\Lambda$ $=$ 10 \textmu m and aperture size of 1x2 mm$^{2}$. The crystal is pumped by a fiber-coupled continuous-wave diode laser (Surelock, Coherent) at 405 nm. A half-wave plate (H) is used to set the polarization of the laser and a plano-convex lens (L$_{1}$) of 300 mm is used to focus the pump beam into the center of the crystal with beam waist \textit{w}\textsubscript{0} $=$ 42.5 \textmu m.

	PPKTP crystal is housed in an oven and its temperature is maintained at 39 \protect\raisebox{0.5pt}{$^\text{o}$}C to obtain degenerate photon pairs at 810 nm. We used a bandpass interference filter (IF) at 810 nm center wavelength with a bandwidth of 10 nm FWHM for collecting the SPDC photons from the residual pump light. The wavelength of the down-converted photons is confirmed using a spectrometer (QEPro, Ocean Insight). The generated orthogonally polarized photon pairs ($|H\rangle$, $|V\rangle$) are collimated using a plano-convex lens (L$_{2}$) of 35 mm and separated using a polarization beam splitter (PBS). Both $|H\rangle$ (idler) and $|V\rangle$ (signal) polarized single photons are coupled into single-mode optical fibers using appropriate collection optics. It should be noted that idler photons are used for heralding and signal photons are passed through HBT interferometer as they are directed to a non-polarizing 50:50 beam splitter (BS). Two outputs of the BS are connected to two fiber coupled detectors, single photon counting modules, SPCMs (A \& B) (SPCM-800-44-FC, Excelitas), which are connected to time-correlated single-photon counter (Time Tagger, Swabian instruments). The idler photons may or may not be detected using another SPCM (C), leading to two different measurement configurations as shown in Fig.\,\ref{Fig1}\tr{(b),(c)}. In the first configuration as illustrated in Fig.\,\ref{Fig1}\tr{(b)}, idler photons are not detected and the purity of single photons is measured for unheralded condition, $g_{\mathrm{uh}}^{(2)}(\tau)$, using only two detectors. In the second configuration as illustrated in Fig.\,\ref{Fig1}\tr{(c)}, all three detectors are used for heralded second-order correlation measurement $g_{\mathrm{h}}^{(2)}(\tau)$. These two configurations are used to investigate the purity of single photons as a function of time bin and different pump powers.

	\subsection{Experimental results and characterization}

	Using the experimentally obtained photon arrival time data from the SPDC process for different pump power,  $g^{(2)}(\tau)$ is calculated for unheralded and heralded single-photon source using the counts described in Eq.\,\eqref{g2uherald} and Eq.\,\eqref{g2herald}.

	\begin{figure}[h!]
		\begin{center}
			\includegraphics[width=0.46\textwidth]{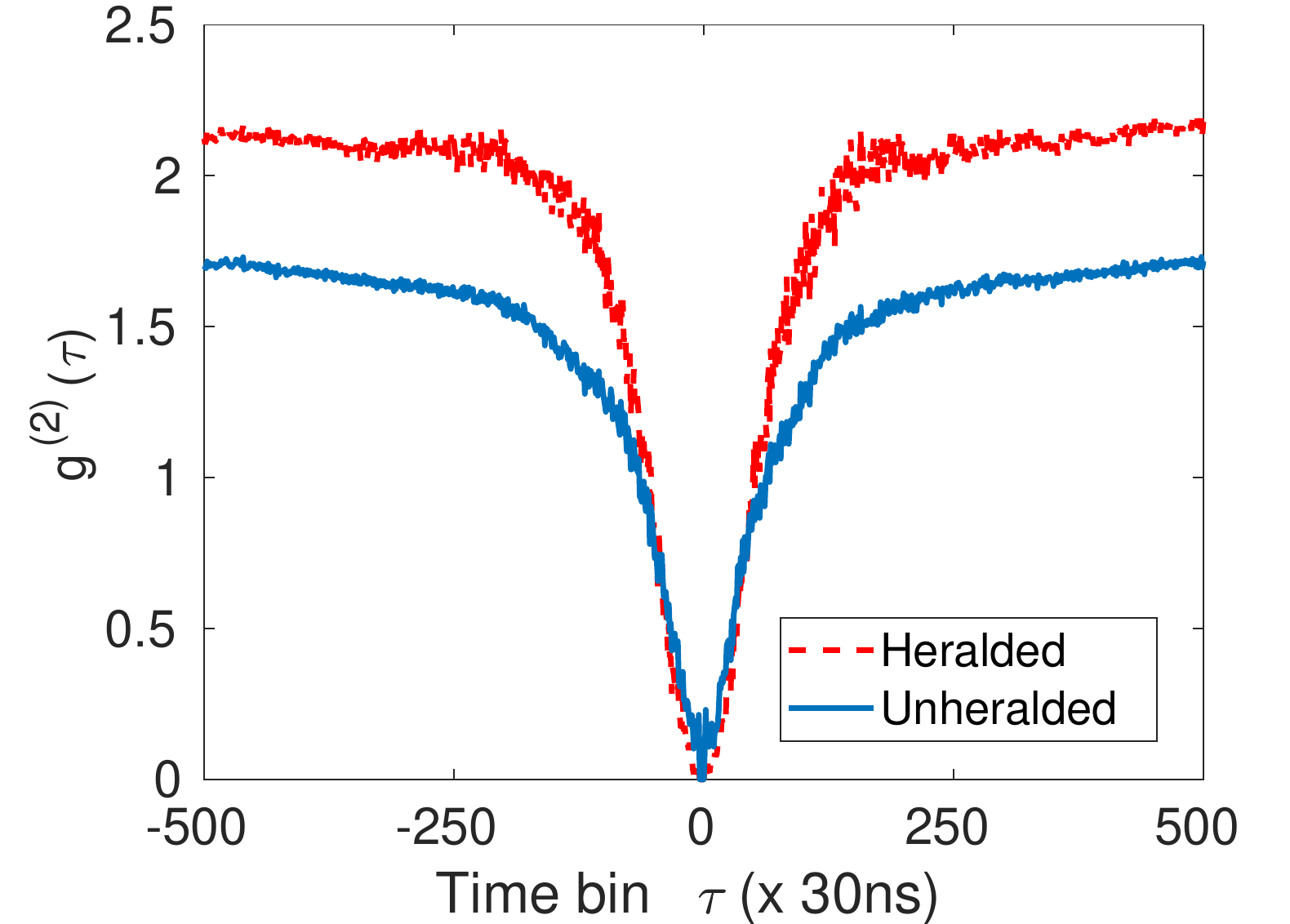}
		\end{center}
		\caption{Variation in $g^{(2)}(\tau)$ as a function of $\tau$, together for unheralded and heralded conditions when photons were recorded at the pump power of 30 mW using single-mode fibers. We can see that $g^{(2)}(0) = 0$ in both the cases.}  
		\label{g2UnHandHcomp}
	\end{figure}
		\begin{figure}[h!]
		\centering
		\includegraphics[width=0.46\textwidth]{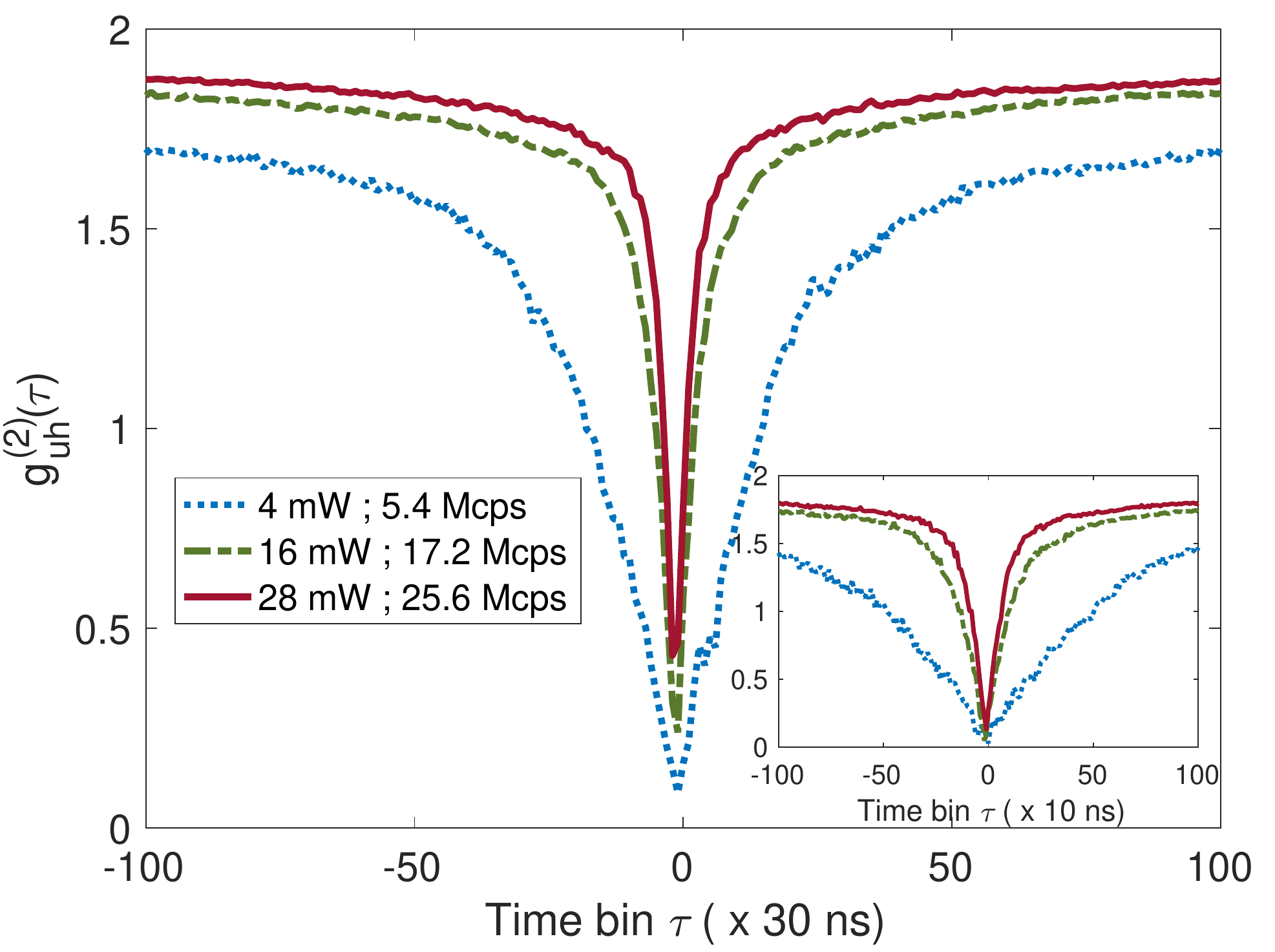}
		\caption{$g_{\mathrm{uh}}^{(2)}(\tau)$ as a function of time bin for minimum $\tau$ set at 30 ns. Function $g_{\mathrm{uh}}^{(2)}(\tau)$ is calculated, for photon counts recorded using multi-mode optical fibers to the detectors, for 200 samples for each $\tau$  and is  averaged over the total number of bins with at least one-photon detection. We can see a clear dip indicating the presence of single photons when smaller time bin is chosen. Value of $g_{\mathrm{uh}}^{(2)}(0)$ for 4 mW, 16 mW and 28 mW pump power is 0.08, 0.23, and 0.43, respectively. Inset: $g_{\mathrm{uh}}^{(2)}(\tau)$ is calculated for $\tau$ set to 10 ns for same three pump power and the value is 0.03, 0.048 and 0.13 for the three pump powers, respectively.}
		\label{g2multimode}
	\end{figure} 
\begin{figure}[h!]
	\centering
	\begin{subfigure}[ ]{ 
		\includegraphics[width=0.4\textwidth]{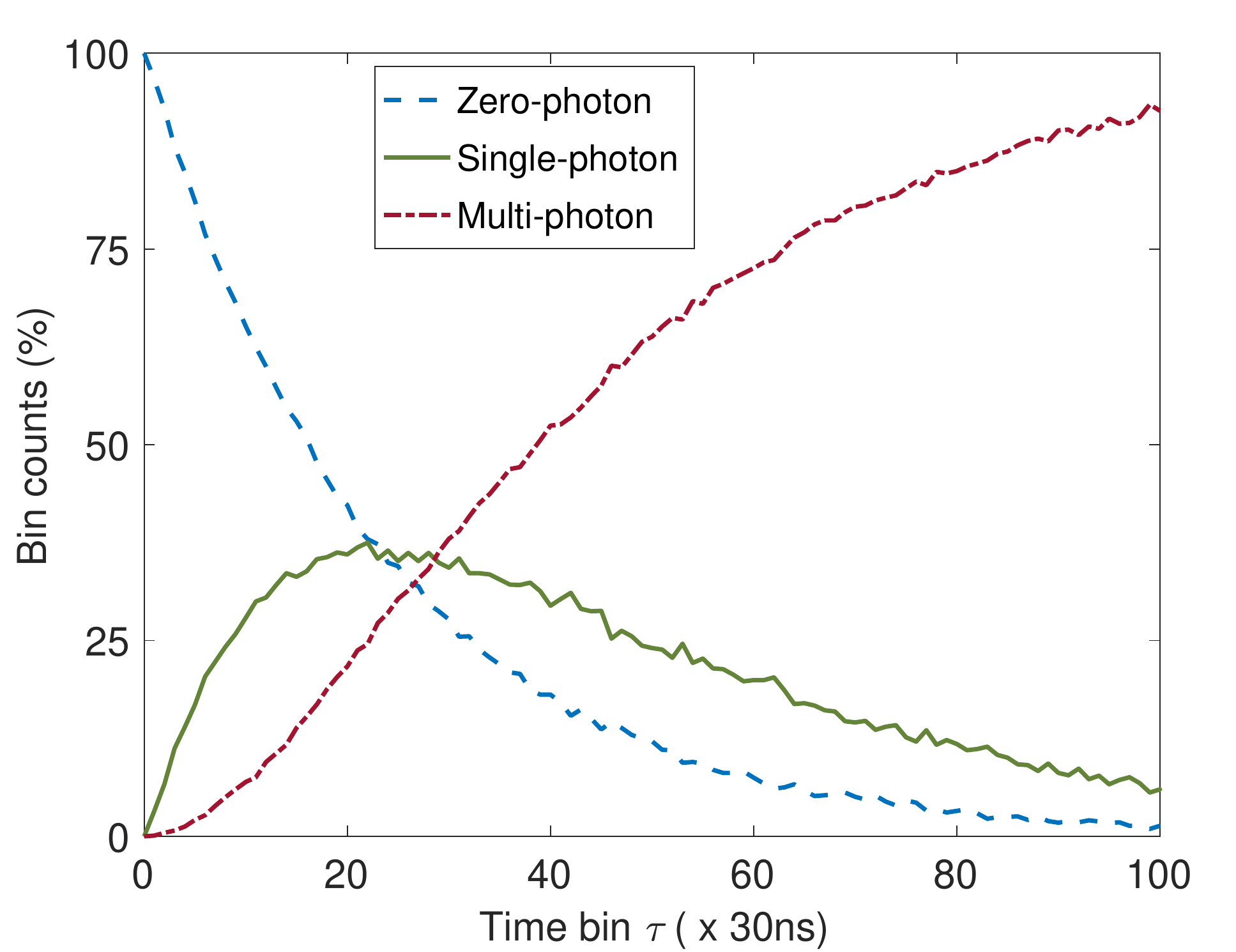}
			\label{5a}}
	 \end{subfigure}
		\begin{subfigure}[ ]{ 
		\includegraphics[width=0.4\textwidth]{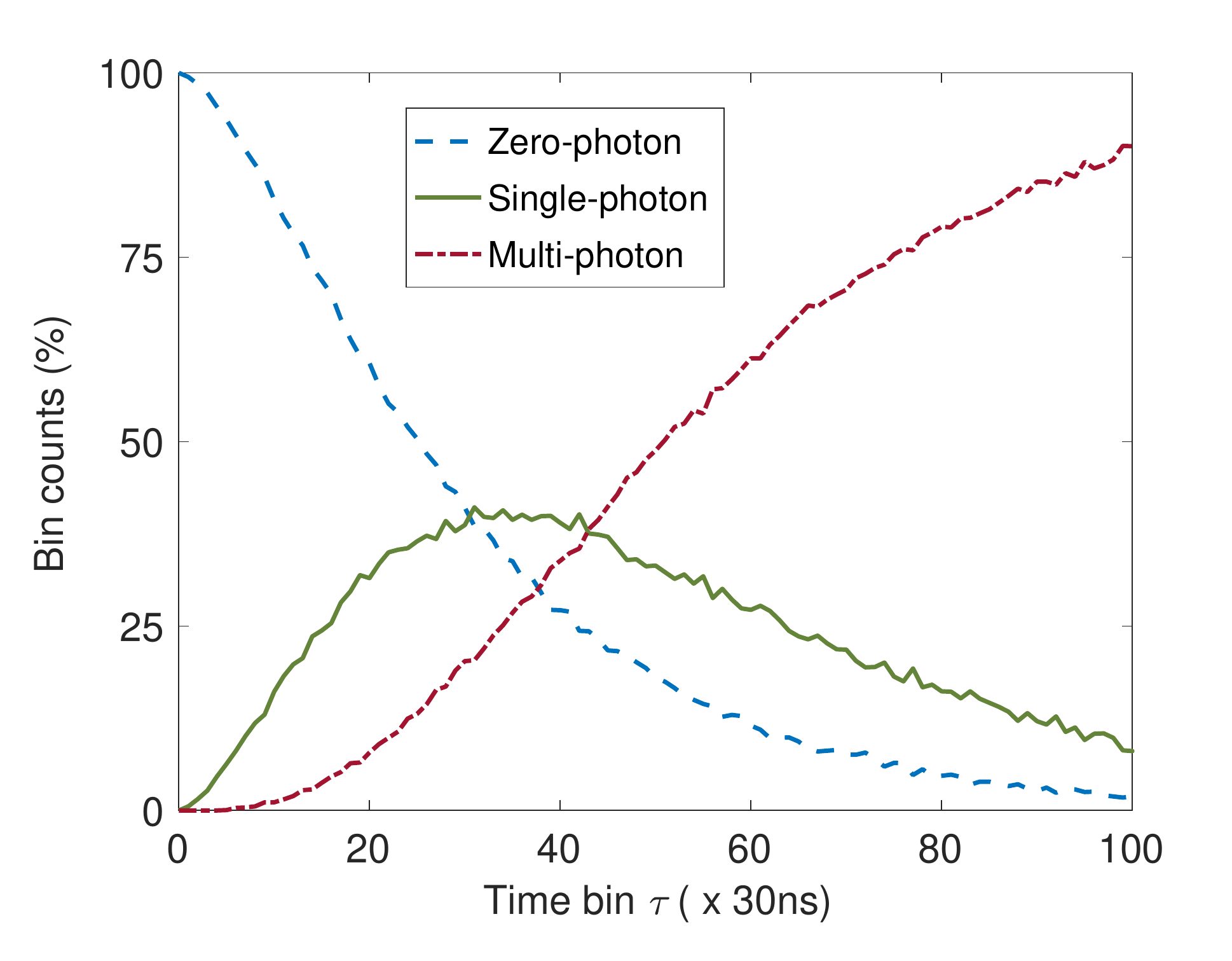}
		\label{5b}}
        \end{subfigure}
\begin{subfigure}[ ]{ 
		\includegraphics[width=0.4\textwidth]{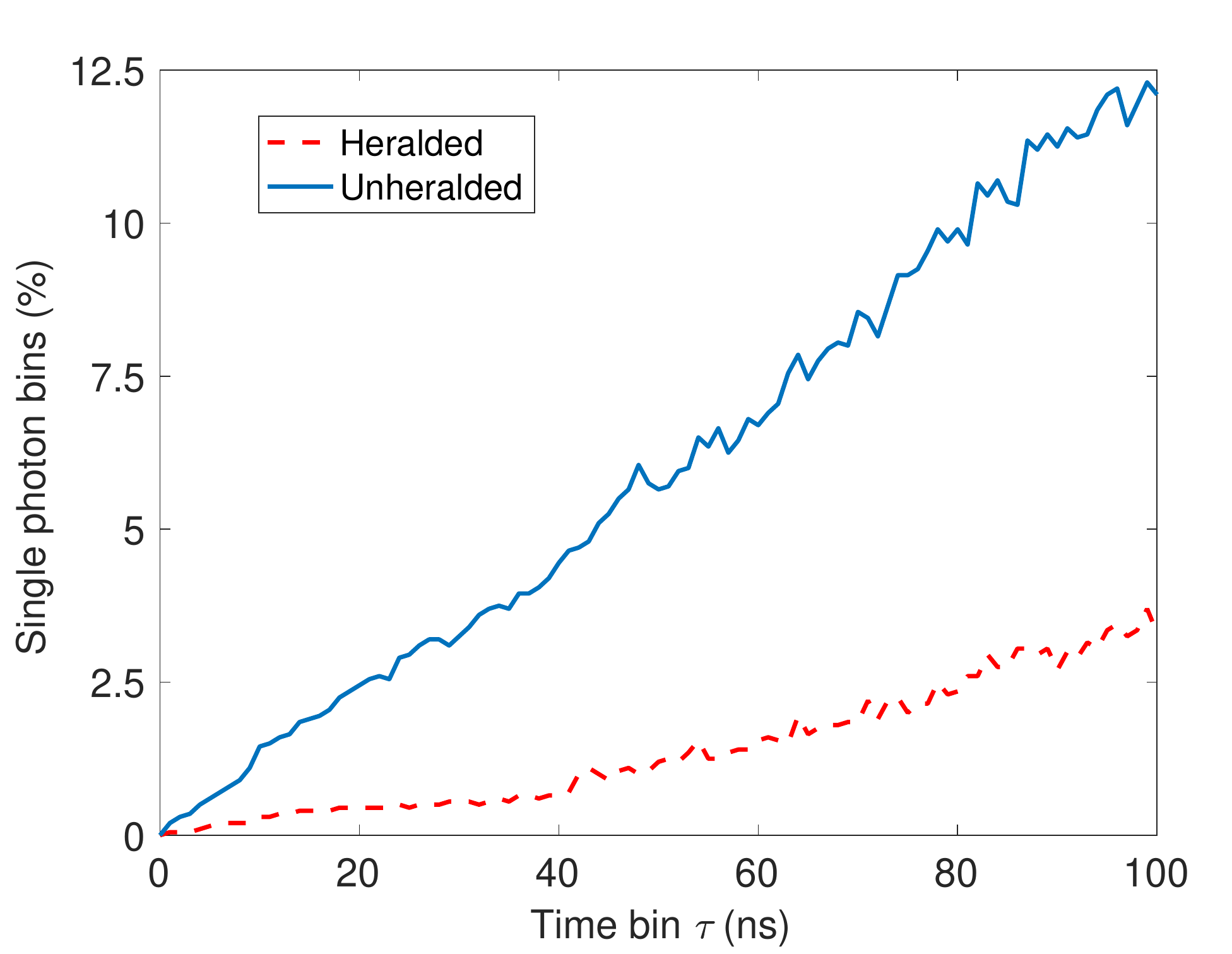}
		\label{5c}}
        \end{subfigure}
	\caption{Variation of percentage of bins with single-photon, multi-photon and no-photon with increase in time bin $\tau$. (a) Unheralded source (b) Heralded source (c) Comparison of bins with single photons for small $\tau$. When time bin is small and multi-photon detection is almost absent, higher number of single-photon bins are recorded for unheralded source of photons when compared to heralded source. This variation is calculated when pump power was set to 30 mW and SPDC output was connected to SPCM via single-mode fiber.}  
	\label{sp_mp_np}
\end{figure}
\begin{figure}[h!]
	\begin{center}
		\includegraphics[width=0.46\textwidth]{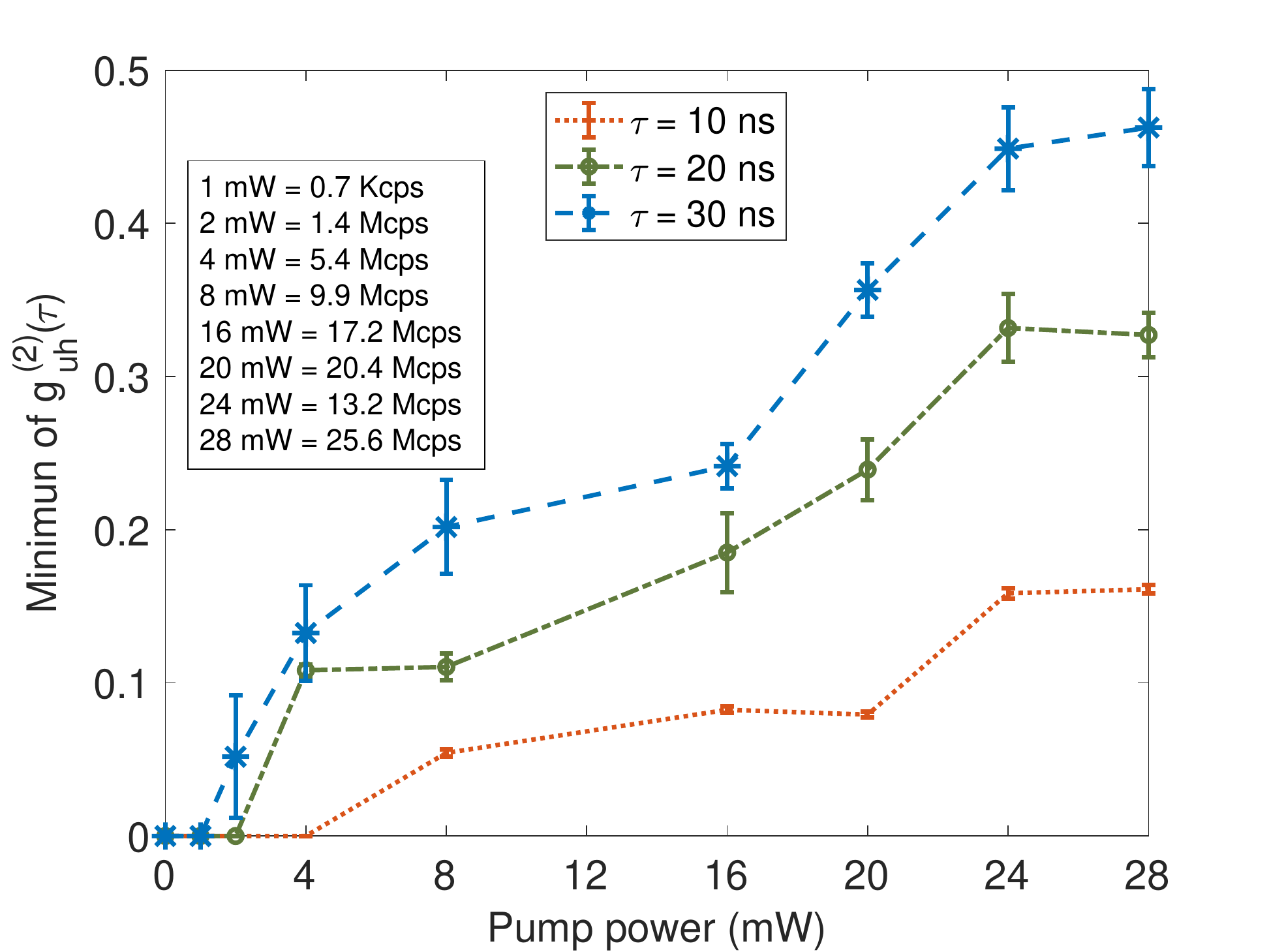}
	\end{center}
	\caption{Variation of minimum value of $g_{\mathrm{uh}}^{(2)}(\tau)$ with increasing pump power when multi-mode fibers were used to connect the SPDC output to the SPCM. Variation of $g_{\mathrm{uh}}^{(2)}(\tau)$ is shown for three $\tau$, 10 ns, 20 ns and 30 ns. With increase in power and bin width $\tau$ we clearly see an increase in minimum value of $g_{\mathrm{uh}}^{(2)}(\tau)$.}  
	\label{g2vsbin}
\end{figure}

	In Fig.\,\ref{g2UnHandH}, the value of $g^{(2)}(\tau)$ is calculated from the photon counts recorded using single-mode optical fibers from SPDC output to the SPCMs is shown with increase in time bin $\tau$ along both the directions from the fixed point in the photon detection time. Fig.\,\ref{2a} shows the value of  $g_{\mathrm{uh}}^{(2)}(\tau)$ calculated for unheralded measurement setup using two SPCMs as illustrated in Fig.\,\ref{Fig1}\tr{(b)}. Fig.\,\ref{2b} shows the value of $g_{\mathrm{h}}^{(2)}(\tau)$ calculated for the heralded source with three SPCM setting as illustrated in Fig.\,\ref{Fig1}\tr{(c)}. Minimum $\tau$ was set to 30 ns $>$ 22 ns (detector dead time) and $g^{(2)}(\tau)$ was calculated for multiples of $30$ ns by averaging over the total number of bins with at least one-photon detection from a sample of 200 bins for each $\tau$. For both, unheralded and heralded source with photon generation rate of upto 1.5 Million counts per second (Mcps) using three different pump power we have shown the corresponding $g^{(2)}(\tau)$ values. We can see a sharp dip leading to $g^{(2)}(0) = 0$, indicating the purity of single photons when smaller time bin is chosen. The value of $g^{(2)}(0)$ can be extrapolated from $g^{(2)}(\tau)$ plots. With increase in pump power, due to increase in number of bins with multi-photon, the number of time bins containing single-photon decreases  resulting in the increase in slope of $g_{\mathrm{uh}}^{(2)}$ with the initial increase in $\tau$. In Fig.\,\ref{g2UnHandHcomp} we show the value of  $g^{(2)}(\tau)$ as a function of $\tau$ together for both, unheralded and heralded configurations when photons were recorded at the rate of $1.5$Mcps (pump power of 30mW with single-mode fiber).  In both cases, for a smaller $\tau$ and with increase in $\tau$  till $g^{(2)}(\tau) =1$  the $g^{(2)}(\tau)$ values are identical. This indicates that the fraction of bins with multi-photon detection are almost identical for both case with increase in $\tau$ until $g^{(2)}(\tau) =1$.

When photon rate is increased by using multi-mode optical fiber connections from SPDC output to the to the SPCMs we can see that even for smaller time bin, $g_{\mathrm{uh}}^{(2)}(\tau) > 0$. In Fig.\,\ref{g2multimode} we show this behaviour for comparable pump powers used in Fig.\,\ref{g2UnHandH}. 
Minimum value of $g_{\mathrm{uh}}^{(2)}$ when minimum of $\tau$ was set to $30 ns$ is 0.08, 0.23, and 0.43 for recorded photon generation rate of 5.4 Mcps, 17.2 Mcps and 25.6 Mcps, respectively. In the inset we show the same when minimum of $\tau$ was set to $10 ns$. The minimum values of $g^{(2)}$ are  0.03, 0.048 and 0.13 for photon generation rate of 5.4 Mcps, 17.2 Mcps and 25.6 Mcps, respectively.

Even though, the low $g^{(2)}(\tau)$ values with smaller time bin indicate the purity of single photons, it does not give us any information about the number of bins with single photons on an average. As sown in Fig.\,\ref{g2UnHandHcomp}, we record identical $g^{(2)}(\tau)$ for single photons using unheralded and heralded settings when the same power is used but the number of bins with single photons may not be same. This can be further analysed by counting the number of bins with single-photon, multi-photon and no-photon.  

We draw a closer look into the way the percentage of bins with single-photon, multi-photon and no-photon  changes with increase in size of time bin $\tau$ for both, unheralded  and heralded  photons as shown in Fig.\,\ref{5a} and Fig.\,\ref{5b}, respectively.  For calculations,  data was collected  when the pump power is set to 30 mW and SPDC output was connected to SPCM via single-mode optical fiber and percentage is calculated from a sample of 1000 bins for each time bin $\tau$.   For both, unheralded and heralded case we clearly see an increase in percentage of bins with multi-photons and decrease in percentage of bins with no-photon.  Even the initial increase in the percentage of bins containing single-photon with increasing $\tau$ looks similar.  However, a closer look shows that the percentage of bins with single photons increases at a faster rate for unheralded case when compared to heralded case.  This is explicitly shown by narrowing down to the smaller $\tau$ in Fig.\,\ref{5c} for which the multi-photon bins are almost zero. On an average we record 4 times higher number of single-photon bins for unheralded source of photons when compared to heralded source.  Along with further increase in percentage of single photon bins we also see increase in percentage of multi-photon resulting in increase of $g^{(2)}(\tau)$ value.   In Fig.\,\ref{g2vsbin} we substantiate this further by calculating the minimum value of  $g^{(2)}(\tau)$  for various pump power and $\tau$ when unheralded SPDC output was connected to SPCM via multi-mode fiber.  The value is calculated for sample of 1000 bins  for each power  and the standard  deviation is obtained by calculating for 100 such sample of 1000 bins each.  An increase in minimum value of $g^{(2)}(\tau)$ with increase in pump power is seen and for a choice of smaller time bin, increase is very small.  This shows that a with better ability to resolve photon arrival in time one can configure and increase purity of unheralded single photons and use such source for practical application where probabilistic single photons in time are sufficient. 


\section{conclusion}
\label{conc}

Though SPDC process is highly probabilistic in nature, it is one of the main sources of single and entangled photons for various quantum information processing tasks and quantum optics experiments. Single photons from SPDC are obtained by heralding one of the photons from the pair of down converted photons. However, for various applications, higher rate of single photons is sufficient even if they are probabilistic in time.  In this work we have proposed the use of unheralded single photons from SPDC for such applications. To quantify the purity of single photons from unheralded SPDC process we have presented a revised expression to calculate second-order temporal correlation function $g^{(2)}(\tau)$. The revised approach rely on probabilities of detection in each detector for a fixed time  bin and calculating  $g^{(2)}(\tau)$ for the selected time bin $\tau$ rather than the usually followed method of determining probabilities from the full temporal behaviour and averaging for the time bin.  Since most of the quantum operations and experiments are performed in a smaller time windows rather than on a full temporal behaviour of the quantum state, this method seems to be more appropriate to calculate  $g^{(2)}(\tau)$ for photons from probabilistic source like SPDC.  By calculating  $g^{(2)}(\tau)$  as a function of time bin $\tau$ for different pump power, we have shown the value of  $\tau$ for which $g^{(2)}(\tau) = 0$ gets narrower with increase in pump power and becomes greater than 0 for very high pump power.  

Our characterization of percentage of bins with single-photon, multi-photon and no-photon with increase in $\tau$ for heralded and unheralded detection has shown that the rate of single-photon generation is higher for unheralded case when the number of bins with multi-photon is negligible. This also helps us to associate the combination of time bin and pump power with the purity of single photons and choose the appropriate time bin and pump power accordingly to keep the multi-photon noise to the minimum for both, heralded and unheralded setting. In summary, with a combination of pump power, choice of $\tau$, better detector efficiency (resolving and detection) with low dead time and filtering of pump power from entering detector we can ensure detection of unheralded single photons with high purity from SPDC process. It can be effectively used for applications where gated or on-demand sources of single photons are not essential.


\vskip 0.2in

{\bf Acknowledgement: } We thank R. S. Gayatri and Abhaya S Hegde for their inputs and useful discussion. We acknowledge the support from the Office of Principal Scientific Advisor to Government of India, project no. Prn.SA/QSim/2020.



\begin{thebibliography}{10}
	
	\bibitem{BPM97}
	D. Bouwmeester, J.-W. Pan, K. Mattle, M. Eibl, H. Weinfurter, and A. Zeilinger, Experimental quantum teleportation,
	\href{https://doi.org/10.1038/37539}{Nature \textbf{390}, 575 (1997).}
	
	\bibitem{MWK96}
	K. Mattle, H. Weinfurter, P. G. Kwiat, and A. Zeilinger, Dense Coding in Experimental Quantum Communication, 
	\href{https://doi.org/10.1103/PhysRevLett.76.4656}{Phys. Rev. Lett. \textbf{76}, 4656 (1996).}
	
	\bibitem{JSW00}
	T. Jennewein, C. Simon, G. Weihs, H. Weinfurter, and A. Zeilinger, Quantum Cryptography with Entangled Photons, 
	\href{https://doi.org/10.1103/PhysRevLett.84.4729}{Phys. Rev. Lett. \textbf{84}, 4729 (2000).}
	
	\bibitem{PBW98}
	J.-W. Pan, D. Bouwmeester, H. Weinfurter, and A. Zeilinger, Experimental Entanglement Swapping: Entangling Photons That Never Interacted, 
	\href{https://doi.org/10.1103/PhysRevLett.80.3891}{Phys. Rev. Lett. \textbf{80}, 3891 (1998).}
	
		\bibitem{KLM01}
	E. Knill, R. Laflamme and G. J. Milburn, A scheme for efficient quantum computation with linear optics,
	\href{https://doi.org/10.1038/35051009}{Nature \textbf{409}, 46 (2001).}

	\bibitem{MZK12}
	X.-s. Ma, S. Zotter, J. Kofler, R. Ursin, T. Jennewein, C. Brukner, and A. Zeilinger, Experimental delayed-choice entanglement swapping,
	\href{https://doi.org/10.1038/nphys2294}{Nature Phys \textbf{8}, 479 (2012).}
	
	\bibitem{APS21} 
	A. Anwar,  C. Perumangatt, F. Steinlechner, T. Jennewein, and  A. Ling, Entangled photon-pair sources based on three-wave mixing in bulk crystals,
	\href{https://aip.scitation.org/doi/10.1063/5.0023103}{Rev. Sci. Instrum. \textbf{92}, 041101 (2021).}
	
	\bibitem{KE07} K. Edamatsu, Entangled Photons: Generation, Observation, and Characterization, \href{https://doi.org/10.1143/JJAP.46.7175}{Jpn. J. Appl. Phys. \textbf{46}, 7175 (2007).}
	
	\bibitem{KMW95}
	P. G. Kwiat, K. Mattle, H. Weinfurter, A. Zeilinger, A. V .Sergienko, and Y.Shih, New High-Intensity Source of Polarization-Entangled Photon Pairs,
	\href{https://doi.org/10.1103/PhysRevLett.75.4337}{Phys. Rev. Lett. \textbf{75}, 4337 (1995).}
	
	\bibitem{KFM04}
	C. E. Kuklewicz, M. Fiorentino, G. Messin, F. N. C. Wong, and J. H. Shapiro, High-flux source of polarization-entangled photons from a periodically poled KTiOPO${_4}$ parametric down-converter,
	\href{https://doi.org/10.1103/PhysRevA.69.013807}{Phys. Rev. A \textbf{69}, 013807 (2004).}
	
	\bibitem{FKW05}
	M. Fiorentino, C. E. Kuklewicz and F. N. C. Wong, Source of polarization entanglement in a single periodically poled KTiOPO${_4}$ crystal with overlapping emission cones,
	\href{https://doi.org/10.1364/OPEX.13.000127}{Opt. Express \textbf{13}, 127 (2005) .}
	
	\bibitem {BCR09} E. Bocquillon, C. Couteau, M. Razavi, R. Laflamme, and G. Weihs, Coherence measures for heralded single-photon sources,
	\href{https://journals.aps.org/pra/abstract/10.1103/PhysRevA.79.035801}{Phys. Rev. A \textbf{79}, 035801 (2009).}
	
	\bibitem{FAT04} 
	S. Fasel, O. Alibart, S. Tanzilli, P. Baldi, A. Beveratos, N. Gisin and H. Zbinden, High-quality asynchronous heralded single-photon source at telecom wavelength,
	\href{https://iopscience.iop.org/article/10.1088/1367-2630/6/1/163}{New J. Phys. \textbf{6}, 163 (2004).} 
	
	\bibitem{EFM11}
	M. D. Eisaman, J. Fan, A. Migdall, and S. V. Polyakov, Invited Review Article: Single-photon sources and detectors, 
	\href{https://aip.scitation.org/doi/10.1063/1.3610677}{Rev. Sci. Instrum. \textbf{82}, 071101 (2011).}
	
	\bibitem{RHH09}
	R. H. Hadfield, Single-photon detectors for optical quantum information applications, 
	\href{https://www.nature.com/articles/nphoton.2009.230}{Nature Photon \textbf{3}, 696 (2009).}
	
	\bibitem{AC2012}
	A. Christ and C. Silberhorn, Limits on the deterministic creation of pure
	single-photon states using parametric down-conversion, 
	\href{https://doi.org/10.1103/PhysRevA.85.023829}{Phys. Rev. A \textbf{85}, 023829 (2012).}
	
	\bibitem{EM2020}
	E. Meyer-Scott, C. Silberhorn, and A. Migdall, Single-photon sources: Approaching the ideal through multiplexing, \href{https://doi.org/10.1063/5.0003320}{Rev. Sci. Instrum. \textbf{91}, 041101 (2020).}
	
	\bibitem{PM2008} 
	P. J. Mosley, J. S. Lundeen, B. J. Smith, P. Wasylczyk, A. B. U'Ren, C. Silberhorn, and I. A. Walmsley, Heralded generation of ultrafast single photons in pure quantum states. \href{https://doi.org/10.1103/PhysRevLett.100.133601} {Phys. Rev. Lett. \textbf{100}, 133601 (2008).}
	
	\bibitem{AZ2011} 
	X. Ma, S. Zotter, J. Kofler, T. Jennewein, and A. Zeilinger, Experimental generation of single photons via active multiplexing, \href{https://doi.org/10.1103/PhysRevA.83.043814} {Phys. Rev. A \textbf{83}, 043814 (2011).}
	
	\bibitem{WT2017} 
	M. G. Puigibert G. H. Aguilar, Q. Zhou, F. Marsili, M. D. Shaw, V. B. Verma, S. W. Nam, D. Oblak, and W. Tittel, Heralded Single Photons Based on Spectral Multiplexing and Feed-Forward Control,
	\href{https://doi.org/10.1103/PhysRevLett.119.083601}{Phys. Rev. Lett. \textbf{119}, 083601 (2017).}
	
	\bibitem{SKB09}
	M. Scholz, L. Koch, and O. Benson, Statistics of Narrow-Band Single Photons for Quantum Memories Generated by Ultrabright Cavity-Enhanced Parametric Down-Conversion,
	\href{https://journals.aps.org/prl/abstract/10.1103/PhysRevLett.102.063603}{Phys. Rev. Lett. \textbf{102}, 063603 (2009).}
	
	
	\bibitem{MMM19} M. Massaro, E. Meyer-Scott, N. Montaut, H. Herrmann and C. Silberhorn, Improving SPDC single-photon sources via extended heralding and feed-forward control, \href{https://doi.org/10.1088/1367-2630/ab1ec3}{New J. Phys. {\bf 21} 053038 (2019).}
	
	\bibitem{BT56}
	R. H. Brown and R. Q. Twiss, Correlation between photons in two coherent beams of light, 
	\href{https://doi.org/10.1038/177027a0}{Nature \textbf{177}, 27 (1956).}
	
	\bibitem{GRA86} 
	P. Gragier, G. Roger and A. Aspect,  Experimental Evidence for a Photon Anticorrelation Effect on a Beam Splitter: A New Light on Single-Photon Interferences,
	\href{http://iopscience.iop.org/0295-5075/1/4/004}{Europhys. Lett. \textbf{1}, 173 (1986).}
	
	
	\bibitem{MW1994} 
	L. Mandel and E. Wolf, Optical Coherence and Quantum Optics, Cambridge: Cambridge University Press (1994).
	
	\bibitem{SGA21} K. Muhammed Shafi, R. S. Gayatri, A. Padhye, and C. M. Chandrashekar, Bell-inequality in path-entangled single photon and purity test, \href{https://arxiv.org/abs/2112.05039}{arXiv:2112.05039 (2021).}
	
	
	
	\bibitem{SCH22}  K. Muhammed Shafi, Prateek Chawla, Abhaya S. Hegde, R. S.  Gayatri, A.  Padhye,  and C. M. Chandrashekar,  Multi-bit quantum random number generator from path-entangled single photons,   \href{https://arxiv.org/abs/2202.10933}{arXiv:2202.10933  (2022).}
	
	
	\bibitem{Bec07} 
	M. Beck, Comparing measurements of $g^{(2)}$(0) performed with different coincidence detection techniques
	\href{https://opg.optica.org/josab/fulltext.cfm?uri=josab-24-12-2972&id=144812}{J. Opt. Soc. Am. B \textbf{24}, 2972 (2007).}
	
	\bibitem{JFC18} 
	C. Joshi, A. Farsi, S. Clemmen, S. Ramelow \& A. Gaeta, Frequency multiplexing for quasi-deterministic heralded single-photon sources
	\href{https://www.nature.com/articles/s41467-018-03254-4}{Nature Communications, \textbf{9}, 847 (2018).}
	
	
	
	
	
\end{thebibliography}
\end{document}